\documentclass[twocolumn,showpacs,amsmath,amssymb,amsfonts,superscriptaddress,floatfix,prl,aps]{revtex4-1}
\usepackage{graphicx}
\usepackage{color}
\usepackage{ulem}

\def\bea{\begin{eqnarray}}
\def\eea{\end{eqnarray}}
\def\ben{\begin{equation}}
\def\een{\end{equation}}
\def\benu{\begin{enumerate}}
\def\enu{\end{enumerate}}
\def\bei{\begin{itemize}}
\def\eei{\end{itemize}}

\def\n{n}

\def\sss{\scriptscriptstyle\rm}

\def\1var{(\bx_1...\bx\N)}

\def\bp{{\bf p}}
\def\br{{\bf r}}

\def\bx{{\br t}}

\def\s{_{\sss S}}
\def\xc{_{\sss XC}}

\def\N{_{\sss N}}
\def\H{_{\sss H}}

\def\ext{_{\rm ext}}

\def\LDA{^{\rm LDA}}

\def\ee{_{\rm ee}}

\def\sph_int{ {\int d^3 r}}

\newcommand{\nn}{\nonumber}
\def\bA{{\bf A}}
\def\bS{{\bf S}}
\def\bB{{\bf B}}
\def\bL{{\bf L}}
\def\bM{{\bf M}}
\def\bm{{\bf m}}

\begin{document}

\normalem 

\title{Demonstration of Optimal Control of Laser Induced Spin-Orbit Mediated Ultrafast Demagnetization}
\author{P. Elliott}
\email{pelliott@mpi-halle.mpg.de}
\affiliation{Max-Planck-Institut f\"ur Mikrostrukturphysik, Weinberg 2, D-06120 Halle, Germany.}
\author{K. Krieger}
\affiliation{Max-Planck-Institut f\"ur Mikrostrukturphysik, Weinberg 2, D-06120 Halle, Germany.}
\author{J. K. Dewhurst}
\affiliation{Max-Planck-Institut f\"ur Mikrostrukturphysik, Weinberg 2, D-06120 Halle, Germany.}
\author{S. Sharma}
\affiliation{Max-Planck-Institut f\"ur Mikrostrukturphysik, Weinberg 2, D-06120 Halle, Germany.}
\affiliation{Department of physics, Indian Institute for technology-Roorkee, 247667 Uttarkhand, India}
\author{E. K. U. Gross}
\affiliation{Max-Planck-Institut f\"ur Mikrostrukturphysik, Weinberg 2, D-06120 Halle, Germany.}
\date{\today}

\begin{abstract}
Laser induced ultrafast demagnetization is the process whereby the magnetic moment of a ferromagnetic material is seen to drop significantly on a timescale of $10-100$s of femtoseconds due to the application of a strong laser pulse. If this phenomenon can be harnessed for future technology, it offers the possibility for devices operating at speeds several orders of magnitude faster than at present. A key component to successful transfer of such a process to technology is the controllability of the process, i.e. that it can be tuned in order to overcome the practical and physical limitations imposed on the system. In this paper, we demonstrate that the spin-orbit mediated form of ultrafast demagnetization recently investigated [arXiv:1406.6607] by ab-initio time-dependent density functional theory (TDDFT) can be controlled. To do so we use quantum optimal control theory (OCT) to couple our TDDDT simulations to the optimization machinery of OCT. We show that a laser pulse can be found which maximizes the loss of moment within a given time interval while subject to several practical and physical constraints. Furthermore we also include a constraint on the fluence of the laser pulses and find the optimal pulse that combines significant demagnetization with a desire for less powerful pulses. These calculations demonstrate optimal control is possible for spin-orbit mediated ultrafast demagnetization and lay the foundation for future optimizations/simulations which can incorporate even more constraints.
\end{abstract}

\maketitle

\section{Introduction}

Control of quantum dynamics using tailored laser pulses is a long standing goal of modern physics\cite{KRGT89,PDR88,HTC83,RVMK00,CK07,S14,XZL13} as it opens up a whole new world of possibilities for future technologies. Faster, smaller, and more efficient devices could be constructed if we could master control over the charge and spin dynamics of electrons on the nanoscale\cite{KKR10}. However precisely at these very short length and time scales, quantum effects are strong, which makes it difficult to exert this control. With the advent\cite{WLPW92} of laser pulse shapers that can tailor the laser field to a given shape, there was now a tool that could be used for control of quantum dynamics. The challenge is finding the shape of the laser pulse that produces the desired dynamics.

Optimal control theory (OCT) is a method developed\cite{BH65,L79} in both Mathematics and Engineering to solve the problem of finding a particular control variable that gives a desired outcome. In our case, we will search for the electric field $\epsilon(t)$ of a laser pulse to control the properties of our system. In general OCT works by creating a target functional of the control field calculated from simulation of the system. Then any constraints on the system are incorporated using penalty functionals, before extremizing the total functional to find the optimal field. OCT can be extended to the realm of quantum mechanics by constructing the target functional using observables given by the time-dependent schr{\"o}dinger equation (TDSE).

For more than a handful of electrons, propagating the TDSE is a computationally intractable problem due to the coulomb interaction between electrons and an alternative approach must be used. Time-dependent density functional theory (TDDFT) is one such approach, which works by mapping the problem to a non-interacting system\cite{RG84}, referred to as the Kohn-Sham (KS) system. This system is defined such that propagating electrons in this system will reproduce the same time dependent density (the probability to find an electron at any given point) as propagating in the exact system using the TDSE. As the KS system is non-interacting, the problem is now computationally tractable. Although, in principle, this mapping is exact, in practice an approximation must be used. The most commonly used approximation, adiabatic local density approximation (ALDA), has been shown to successfully predict absorption spectra of a large range of atoms, molecules, and solids\cite{EFB09,TDDFTbook12,C11}. Thus TDDFT is an outstanding candidate to couple to OCT\cite{WG07} in order to predict laser pulses
for control of quantum dynamics, and has been used successfully for control of charge transfer\cite{RCWR08}, HHG\cite{SBSC14}, strong-field ionization\cite{HRG13,SMR15}, bond-breaking\cite{KCG11}, among others.

Laser-induced ultrafast demagnetization was first observed in the mid $1990$s, whereby a strong femtosecond laser pulse caused a significant loss of the magnetic moment of a thin film of Ni in a time less than 1ps\cite{BMDB96}. Since then, this phenomena has been the subject of much experimental\cite{AAGR84,VBGL92,HMKB97,SBJE97,ABPW97,HGDK99,RVK00,GBB02,SPWD05,MRWP11,SAMC12} and theoretical\cite{ZHLB09,CBO11,IHF13,KDES14} endeavor and several mechanisms have been proposed to explain the demagnetization. In Ref. \onlinecite{KDES14}, ab-initio TDDFT simulations were performed to investigate the demagnetization and found that when spin-orbit interaction was included in the system Hamiltonian, a loss of moment was observed for very short ($5$fs), very intense ($1\times 10^{15}$ W/cm2) laser pulses.  It is this system we wish to control by varying the intensity and frequency of the laser pulse, subject to several practical constraints, in order to maximize the total loss of moment. To do so we utilize the framework developed in Refs. \onlinecite{WG07,KCG11,CBR15} which combines OCT with quantum simulations of spin dynamics.

\section{Background and Methods}

We begin by briefly reviewing TDDFT and OCT, a more thorough discussion can be found here\cite{WG07}. Then we review the results of Ref. \onlinecite{KDES14} that showed ultrafast demagnetization in bulk ferromagnets (Fe,Co,Ni) for short, strong, laser pulses.

\subsection{TDDFT}

The electronic density is defined as 
\bea
\n(\br,t) &=& N \int d\br_2\mathellipsis d\br_N \ \Psi^*(\br,\br_2,\mathellipsis,\br_N,t) \nn \\ 
&& \times \Psi(\br,\br_2,\mathellipsis,\br_N,t)
\eea
where N is the total number of electrons, $\br$ is the spacial coordinate, $t$ is the time, and $\Psi$ is the wavefunction of the TDSE:
\ben
\label{tdse}
i \frac{\partial}{\partial t}\Psi = \hat{H}\Psi
\een
for Hamiltonian:
\ben
\hat{H} = \hat{T} + \hat{V}\ext + \hat{V}\ee
\een
composed of the kinetic energy, $\hat{T}$, the electron-electron interaction, $\hat{V}\ee$, and the external potential, $\hat{V}\ext$, which includes both the electron-nuclear interaction and the electric fields of any laser pulses. We use atomic units throughout unless otherwise stated. TDDFT is founded upon the Runge-Gross theorem\cite{RG84} which proves a $1-1$ correspondence between the time-dependent density and the time-dependent external potential (up to a time-dependent constant) for any electron-electron interaction. Hence all observables of the system are, in principle, unique functionals of the density. In particular, a non-interacting KS system\cite{KS65} can be defined with a unique KS potential that reproduces the time-dependent density of the interacting system and thus predicts all observables of the true system without requiring the costly propagation of Eq. \ref{tdse}. The TDKS equation is:
\ben
\label{tdks}
i \frac{\partial}{\partial t}\phi_j(\br,t) = \left[ - \frac{\nabla}{2} + v\s(\br,t) \right]\phi_j(\br,t)
\een 
with the total density given by
\ben
n(\br,t) = \sum_{j=1}^N |\phi_j(\br,t)|^2
\een
The KS potential, $v\s(\br,t)$, consists of three pieces:
\ben
v\s(\br,t) = v\ext(\br,t)+v\H(\br,t)+v\xc(\br,t)
\een
where $v\H(\br,t)$ is the usual Hartree potential of the instantaneous density, and $v\xc(\br,t)$ is the exchange correlation (XC) potential and is a functional of the density at all previous times, the interacting initial state, and the non-interacting initial KS state. In practice, it must be approximated, with the most common approximation being the ALDA:
\ben
v\xc[n](\br,t) = v\xc\LDA[n(\br,t)] = \left.\frac{de\xc^{unif}}{dn}\right|_{n=n(\br,t)}
\een
which uses just the instantaneous density inputed into the ground-state DFT LDA XC functional and $e\xc^{unif}(n)$ is the XC energy density of the uniform electron gas. The initial KS state is typically the ground-state found from a DFT calculation.

From this starting point, TDDFT has been extended to include non-collinear magnetism and magnetic fields\cite{CVG01}. For this case, we have a non-interacting Pauli KS Hamiltonian\cite{BH72} which is used to propagate $2$ component spinors, from which the density and magnetization density exactly replicate those of the interacting system. This is the formulation we will use for our simulations. The magnetization density operator may be written as:
\ben
\hat{\bm}(\br) = \hat{\bS} \, \hat{n}(\br)
\een
where $\hat{n}(\br)$ is the density operator and in the two-component spinors propagated in our calculations, $\bS=g/2 \, \boldsymbol\sigma$ where  $\{\sigma_x,\sigma_y,\sigma_z\}$ are the familiar Pauli spin matrices and $g$ is the electronic gyromagnetic ratio. 
For periodic boundary conditions, the total moment is then 
\ben
\bM(t) = \int_\Omega d^3 r \ \bm(\br,t)
\een 
where $\Omega$ is a single unit cell. The KS Hamiltonian for our simulations is:
\bea
\label{Ham}
\hat{H}\s(t)&=&\frac{1}{2}\big(\hat{\bp}+\frac{1}{c}\bA\ext(t)\big)^2 +v\s(\hat{\br},t) \nn \\ 
&&+ \frac{1}{c} \hat{\bS}\cdot\bB\s(\hat{\br},t) +\frac{1}{2c^2} \hat{\bS}\cdot\big( \nabla v\s(\hat{\br},t) \times \hat{\bp} \big)  
\eea
where $\hat{\bp}$ is the momentum operator, $\hat{\bS}$ is the vector spin operator, and $c$ is the speed of light. The laser pulse electric field is written as a vector potential, $\bA\ext(t)$ in the velocity gauge as it allows Bloch's theorem to be utilized. The KS magnetic field is written as $\bB\s(\hat{\br},t)=\bB\ext(t)+\bB\xc(\hat{\br},t)$ where $\bB\ext(t)$ is the magnetic field of the applied electromagnetic field and $\bB\xc(\hat{\br},t)$ is the XC magnetic field. The ALDA can be extended to $\bB\xc$ using the LDA rotation method of K\"ubler\cite{KHSW88}. The final term of Eq. (\ref{Ham}) is the spin-orbit coupling (SOC) term, which can be thought of as the interaction between the spin of an electron and the effective magnetic field caused by relativistic motion thought a scalar potential. In a centrosymmetric potential, this term reduced to the well known $\hat{\bL}\cdot\hat{\bS}$ coupling. Propagation with Hamiltonian Eq. (\ref{Ham}) is implemented in ELK\cite{elk}, an all-electron electronic structure code, which was also used for all ground state and time-dependent calculations.

\subsection{Optimal Control Theory}

The central equation of OCT is the target functional $G[u]$:
\ben
\label{tar}
G[u] = G[\Psi[u],u] = J_1[u] + J_2[u]
\een
where $u$ is the control field and $\Psi[u]$ contains the information on how the system responds to the control field. In Quantum OCT (QOCT), $\Psi[u]$ is then the wavefunction, which is a functional of the control field via the TDSE and from which any system observables to be controlled may be calculated. The target functional is generally separated into two pieces, $J_1[u]$ which contains information on the desired dynamics and $J_2[u]$ which is a penalty function in order to satisfy any constraints on the system or control field. The magnitude of the penalty functional is determined by how strongly a constraint must be satisfied.

Once a relevant target functional has been constructed, the goal of OCT is to extremize it and thus find the optimal control field $u$ to best satisfy the balance between desired dynamics and the constraints. There are many choices for the algorithm to perform this optimization, some are general, such as the Nelder-Mead\cite{NM65} or NEWOAU\cite{P04} algorithms, while some are developed for specific types of problem, e.g. in QOCT the ZBR scheme\cite{ZBR98} adds a time dependent auxiliary wavefunction, which is also propagated in time and the overlap with the true wavefunction used to construct the control field.

For our system, we wish to maximize the loss of magnetic moment in a given time interval [0,T] while including practical and physical constraints on the type of laser pulse. Thus $\epsilon(t)$, the electric field of the laser pulse is the control field and 
\ben
J_1[\epsilon] = \langle \Psi[\epsilon](T) | \hat{\bM}_z | \Psi[\epsilon](T) \rangle = \bM_z(T)
\een 
is the target functional to be minimized, i.e. if we choose the initial magnetization $M_z(0)$ of the ferromagnet to be along the $z$-axis, then minimizing $M_z(T)/M_z(0)$ will maximize the loss of moment. 

The constraints on the electric field are that the pulses satisfy Maxwell's equations (details below) and only certain frequencies are used to construct the pulse. The second constraint is of practical nature, as experimentally, 
pulses containing arbitrary frequencies cannot be constructed and often access to a single frequency (or multiples thereof) is only available. From Maxwell's equations, the following constraints on the electric field must be physically satisfied:
\bea
\int_0^T dt \epsilon(t) = 0 \\
\epsilon(0) = 0 = \epsilon(T) 
\eea
Following Ref. \onlinecite{KCG11}, we can satisfy all these constraints by writing the electric field
\ben
\label{eps}
\epsilon(t) = \sum_{n=1}^{N_\omega} \tilde{\epsilon}_n cos(\omega_n t) 
\een
where $N_\omega$ is the number of frequencies to be used and $\tilde{\epsilon}_n$ are the coefficients to be optimized. It can be seen that this choice automatically satisfies the constraints from Maxwell's equations.
The frequencies used for our demonstration are
\ben
\label{omg}
\omega_n = \frac{2\pi n}{T}
\een

\subsection{Ultrafast Demagnetization}

\begin{figure}[t]
\centerline{\includegraphics[width=\columnwidth,angle=-0]{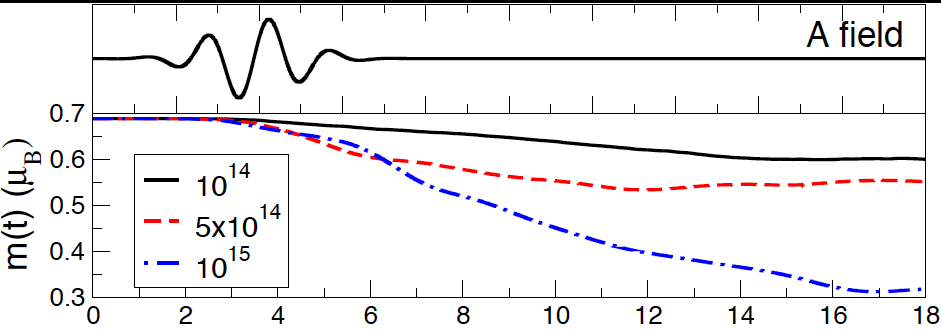}}
\caption{Upper panel: The profile of the laser vector potential (labelled A field).     Lower Panel: The total moment per atom as a function of time for several different peak intensities (given in W/cm2). [Figure reproduced from Ref. \onlinecite{KDES14}.]}
\label{f:ni}
\end{figure}

In Fig. \ref{f:ni}, the results of TDDFT simulations\cite{KDES14} for several laser pulses with differing peak intensities but the same profile are shown. A loss of the total magnetic moment was observed in all cases, with the fraction of moment lost dependent on the field intensity. It was also shown in Ref. \onlinecite{KDES14} that the fraction loss depends on the frequency of the applied laser pulses. Hence the system was a strong candidate for optimal control. The purpose of this paper is test that hypothesis.

It should be pointed out that for less-intense longer-duration pulses and for more realistic system geometries, the required peak intensity and fluence can be reduced by several orders of magnitude\cite{KEDS15}, however the underlying mechanism of demagnetization is the same. Thus we can demonstrate control of this process by focusing on the short strong laser pulses.

\section{Results}

To demonstrate optimal control of the ultrafast demagnetization in Ni, we will attempt to maximize the loss of moment after $T = 600$ au $= 14.4$ fs. We will optimize a pulse of the form given by Eq. (\ref{eps}) using $N_\omega=4$ different laser frequencies defined by Eq. (\ref{omg}). All calculations are performed with a time step of $0.1$ au and $8$x$8$x$8$ $k$-points. For the optimization we choose to use the gradient-free Nelder-Mead simplex algorithm.  To initialize the Nelder-Mead algorithm, $N_\omega+1$ starting points are required, it is instructive to examine these before moving to the results of the optimization.

\subsection{Random Initial Pulses}

\begin{figure}[t]
\centerline{\includegraphics[width=\columnwidth,angle=-0]{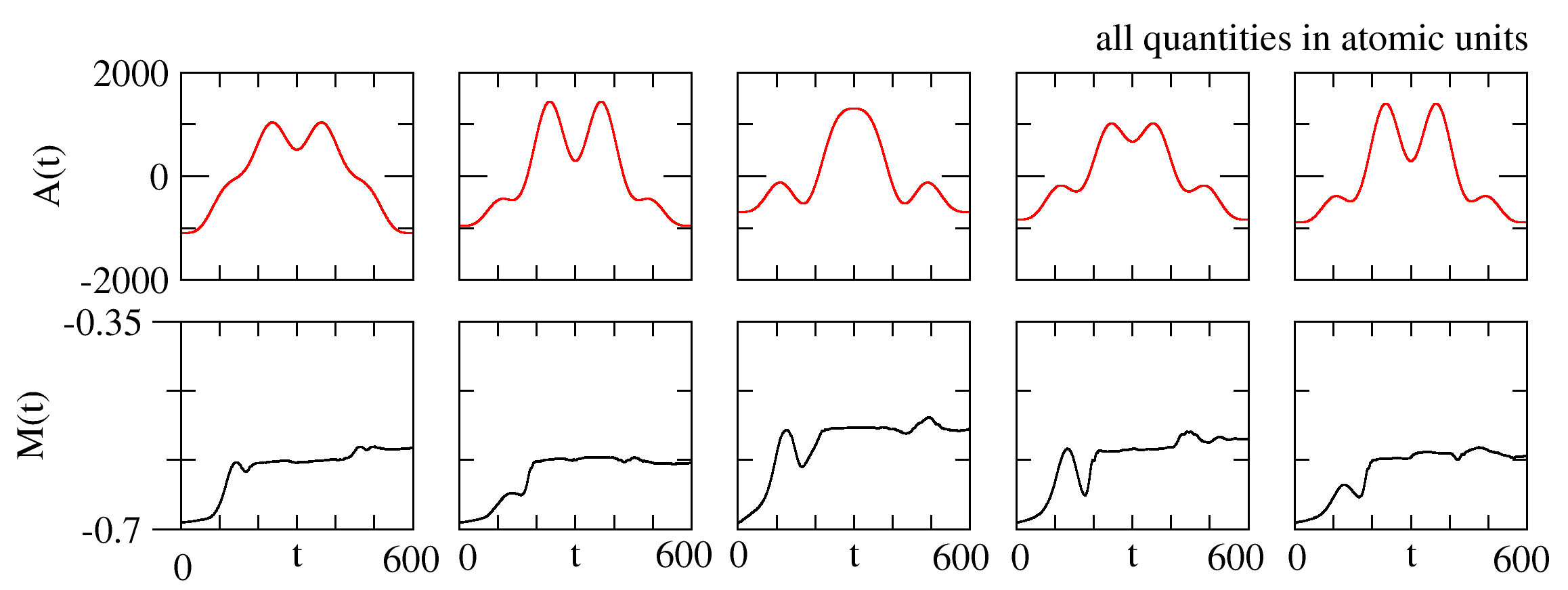}}
\caption{Upper Panels: The vector potentials for initialization laser pulses. Lower Panels: The dynamics of the total moment in the $z$-direction for each pulse.}
\label{f:init}
\end{figure}

To initialize our calculation, we construct $5$ different pulses where the coefficients of Eq. (\ref{eps}) are chosen at random in a suitable range. This range is chosen such that the peak intensity is similar to the demagnetization pulses observed in Ref. \onlinecite{KDES14}. The pulses may be seen in the upper panel of Fig. \ref{f:init}, note that this is the vector potential which can be calculated from the electric field via 
\ben
A(t) = -c\int^t dt' \ \epsilon(t') 
\een
The dynamics of $\bM_z(t)$ are shown in the lower panel and it can be seen that all pulses display demagnetization. If we look at the final time, the average loss of moment is approximately $20\%$. If the optimal control is successful, then this percentage loss should be significantly increased.

\subsection{Maximize Demagnetization}

\begin{figure}[t]
\centerline{\includegraphics[width=\columnwidth,angle=-0]{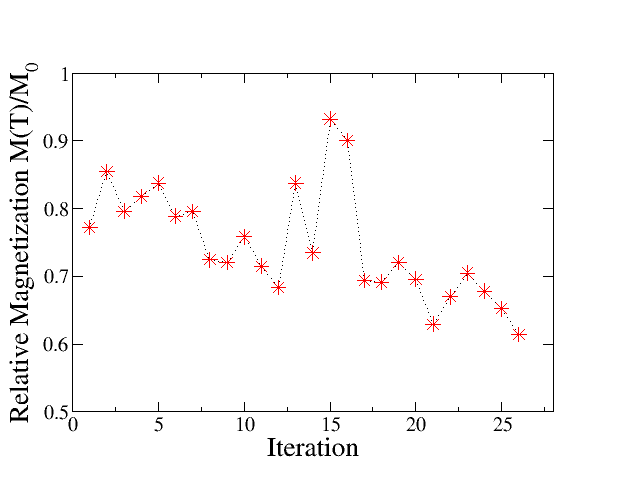}}
\caption{The fraction of magnetic moment loss for each iteration of the Nelder-Mead optimization.}
\label{f:iter}
\end{figure}

From the initial pulses, the Nelder-Mead algorithm then calculates a new set of coefficients from a simple set of rules and then tests how this affects the target functional by performing a TDDFT simulation with the laser field given by these coefficients. It then iterates this procedure and traverses the multidimensional parameter space, searching for the optimal set of coefficients.

In Fig. \ref{f:iter}, we plot the ratio of the final moment after time T to the initial moment for each of the iterations. Although individual iterations can worsen the loss percentage, there is a clear downward trend as better and better pulses are found during the search, indicating that the optimal control is working. Each set of coefficients is a point in the parameter space, at each iteration, the Nelder-Mead algorithm reflects the worst point through the center of mass of the other points. Depending on whether this new point improves upon the next worst point, the algorithm can expand or contract in this direction, otherwise it can reduce all points towards the best point. This explains why individual points may worsen the ratio $\bM_z(T)/\bM_z(0)$.

\begin{figure}[t]
\centerline{\includegraphics[width=\columnwidth,angle=-0]{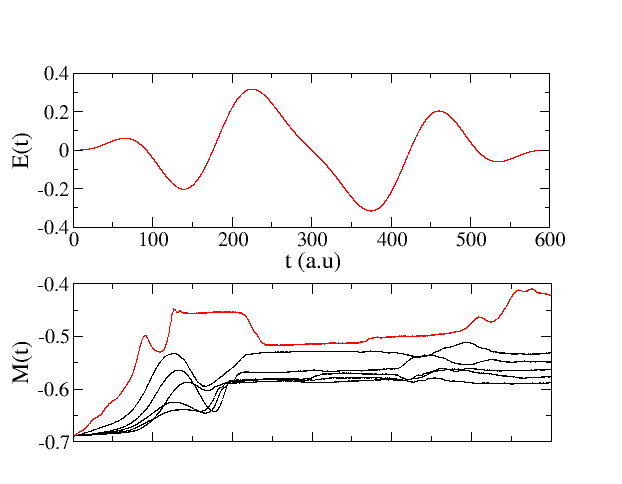}}
\caption{Upper Panel: The optimal laser field found after 30 iteration of the optimal control algorithm. Lower Panel: The magnetization dynamics of $\bM_z$ for this pulse compared to the 5 initialization pulses. }
\label{f:finmag}
\end{figure}

If we look at the result after approximately $30$ iterations, the best pulse the optimal control procedure has found causes a $40\%$ loss of moment. This is twice as good as the random initial pulses used to start the algorithm and is  a clear demonstration that the moment can be successfully controlled using OCT. In Fig. \ref{f:finmag} we show the electric field of this best pulse and also the magnetization dynamics, compared to the initial pulses. Examining the pulse shape compared to the initial pulses, there is no obvious reason why one leads to a larger demagnetization. This is the power of QOCT to find such pulses. We also observe that the magnetization dynamics is a highly non-linear process, in particular for these short pulses, again demonstrating the need for QOCT in order to control the moment.

\subsection{Fluence Constraint}

For many practical reasons, the fluence of the applied pulse should be constrained. The simplest reason is simply efficiency, i.e. using a a pulse with lower energy to achieve the same dynamics as a higher energy pulse. Other reasons include surface damage to the material due to high fluence pulses, heating of the sample (and problems associated with cooling it), or physical restrictions on the laser itself preventing production of high fluence pulses. All of these present significant problems to future technological application, hence we wish to demonstrate how a fluence constraint can be incorporated into our calculations. 

If we add to Eq. (\ref{tar}), the constraint
\ben
\label{flu}
J_2[\epsilon] = \alpha \int_0^T dt \ \epsilon^2(t)
\een
which is proportional to the laser fluence. The free parameter $\alpha$ determines how strong the constraint is, for this calculation we choose $\alpha=0.05$. This parameter was based on examining the results of the previous optimization and choosing $\alpha$ to favor a lower fluence while still maintaining significant demagnetization in the set. 

\begin{figure}[t]
\centerline{\includegraphics[width=\columnwidth,angle=-0]{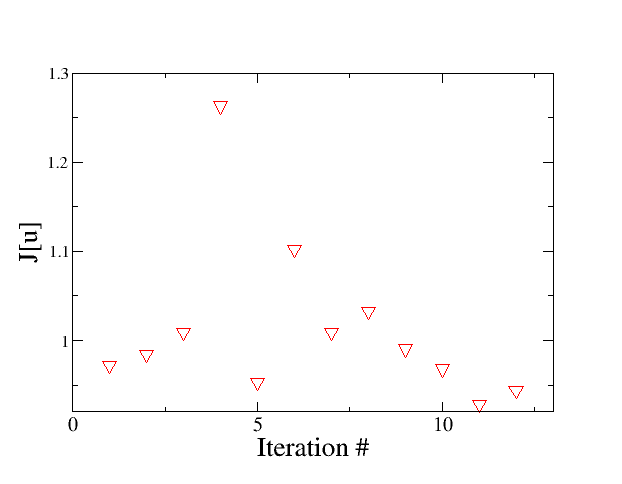}}
\caption{The OCT target functional, $J[u] = M_z[u](T)/M_z(0) + 0.05\times F[u]$ for the electric field $u=\epsilon(t)$ at each iteration of the algorithm, where $F$ is the fluence.}
\label{f:flu}
\end{figure}

In Fig. \ref{f:flu}, we show the value of the total target functional, Eq. (\ref{tar}), for each iteration of the optimization algorithm. Unlike the previous case, we cannot attach a physical meaning to the target functional. so the actual value is not significant, only the trend. Furthermore, when choosing $\alpha$, it was clear that the parameter space is a more complicated environment than the previous case, as a pulse could have the same value of Eq. (\ref{tar}) by either increasing the demagnetization or decreasing the fluence. Due to computational constraints, we stopped the optimization after $12$ iteration, although this was sufficient to see the trend and demonstrate optimal control. We initialize the algorithm using the $5$ best pulses found by calculating the target functional for all the  previous pulses. These are not shown in Fig. \ref{f:flu}, but were part of the optimization search. By using these to start the optimization, we save a large amount of computational time as opposed to using random pulses (although in general this may not be the case if the constraints significantly change the parameter landscape). 

\begin{figure}[ht]
\centerline{\includegraphics[width=\columnwidth,angle=-0]{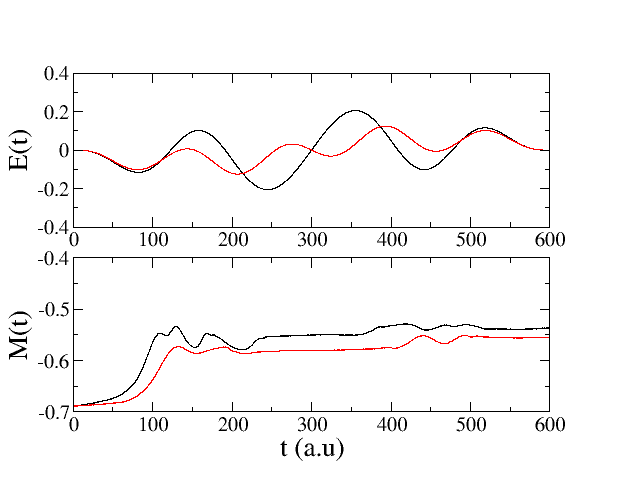}}
\caption{The electric field $E(t)$ (upper panel) and total magnetic moment $\bM_z$ (lower panel) for the reference pulse (solid black line) and the best pulse (dashed red) found during the fluence constraint optimization.}
\label{f:flues}
\end{figure}

To see the power of this optimization, in Fig. \ref{f:flues} we plot the electric fields and the dynamics of $\bM_z(t)$ for two different pulses. These correspond to the best point of Fig. \ref{f:flu} and a reference pulse corresponding to the $G=1.1021$ point. While this point is not the worst point, it was chosen as the final moment is similar to the best point, $-0.537$ and 
$-0.555$ respectively. Thus, the magnetic moments at time $T$ are very close, however the fluences are very different. If we insert the pulses shown in Fig. \ref{f:flues} into the fluence formula given in Eq. \ref{flu}, we find values of $6.437$ a.u for the reference pulse and $2.437$ a.u. for the best pulse. Therefore by using OCT we have reduced the required fluence by $\approx 60\%$.

\section{Conclusions}

In summary we have successfully demonstrated that optimal control of spin-orbit mediated ultrafast demagnetization is possible. For a short time interval, we showed that the loss of moment can be at least doubled (compared to randomly chosen typical pulses) for a system where the available laser frequencies (used to tailor the laser pulse) are constrained. Furthermore we extended the control problem to include a constraint on the laser fluence and demonstrated that QOCT could successfully find a pulse that balances the fluence and demagnetization requirements.
Compared to a reference pulse, this optimal pulse produces almost identical magnetization dynamics, while reducing the fluence by over a factor of $2$. Control of the system is of upmost importance for future technological application (for example, in spintronics), where the desired dynamics and constraints are dictated by 
practical concerns. Any physical phenomenon must be robust to these concerns, and as we have demonstrated, this form of ultrafast demagnetization meets this criteria. Simulation and QOCT of more complicated scenarios, such as longer pulse durations or further constraints on the fluence, intensity, and robustness of the demagnetization, can be build upon this foundation.

\section{Acknowledgements}
This work is supported by SFB 762 of the Deutsche Forschungsgemeinschaft.


\end{document}